\documentclass[supp]{new_tlp}
\usepackage{supp}

\usepackage[utf8]{inputenc}
\usepackage{amsmath,amssymb}
\usepackage{times,helvet,courier}
\usepackage{url}
\usepackage[listofformat=parens,subrefformat=subparens]{subfig}
\usepackage{tikz}
\usetikzlibrary{arrows, shapes}

\usepackage{listings}
\newcommand{\negsubfig}{} 

\newcommand{\clasp}{\textit{clasp}}
\newcommand{\clingcon}{\textit{clingcon}}
\newcommand{\clingo}{\textit{clingo}}

\newcommand{\gringo}{\textit{gringo}}
\newcommand{\iclingo}{\textit{iclingo}}
\newcommand{\oclingo}{\textit{oclingo}}

\newcommand{\dlv}{\textit{dlv}}
\newcommand{\dlvhex}{\textit{dlvhex}}
\newcommand{\acthex}{\textit{acthex}}
\newcommand{\jdlv}{\textit{jdlv}}
\newcommand{\lparse}{\textit{lparse}}

\newcommand{\idp}{\textit{idp}}
\newcommand{\minisat}{\textit{minisat}}
\newcommand{\ZZZ}{\textit{z3}}

\newcommand{\lua}{Lua}
\newcommand{\python}{Python}
\newcommand{\cpp}{C++}
\newcommand{\java}{Java}

\newcommand{\RRG}{\ensuremath{R}}

\title{\textit{Clingo} = ASP + Control: Preliminary Report}

\author[M. Gebser, R. Kaminski, B. Kaufmann, and T. Schaub]{%
  Martin Gebser$^{1,2}$,
  Roland Kaminski$^2$,
  Benjamin Kaufmann$^2$, and
  Torsten Schaub$^2$\thanks{Affiliated with the
                              Simon Fraser University,
                              Burnaby, Canada,
                              and
                              Griffith University,
                              Brisbane, Australia.}
  \\
  $^1$Aalto University, Finland \qquad $^2$University of Potsdam, Germany
  }

\pagerange{\pageref{firstpage}--\pageref{lastpage}}
\volume{[n/a]}
\jdate{[n/a]}
\pubyear{2014}

\pubauthor{Gebser}
\jurl{[n/a]}
\pubdate{[n/a]}

\submitted{[n/a]}
\revised{[n/a]}
\accepted{[n/a]}

\begin{document}

\maketitle

\begin{abstract}
We present the new ASP system \clingo~4.
Unlike its predecessors,
being mere monolithic combinations of the grounder \gringo\ with the solver \clasp,
the new \clingo~4 series offers high-level constructs for realizing complex reasoning processes.
Among others,
such processes feature advanced forms of search, as in optimization or theory solving, or even interact with an environment,
as in robotics or query-answering.
Common to them is that the problem specification evolves during the reasoning process,
either because data or constraints are added, deleted, or replaced.
In fact, \clingo~4 carries out such complex reasoning within a single integrated ASP grounding and solving process.
This avoids redundancies in relaunching grounder and solver programs and benefits from the solver's learning capacities.
\clingo~4 accomplishes this by complementing ASP's declarative input language by control capacities expressed via the embedded scripting languages \lua\ and \python.
On the declarative side, \clingo~4 offers a new directive that allows for structuring logic programs into named and parameterizable subprograms.
The grounding and integration of these subprograms into the solving process is completely modular and fully controllable from the procedural side, viz.\ the scripting languages.
By strictly separating logic and control programs,
\clingo~4 also abolishes the need for dedicated systems for incremental and reactive reasoning, like \iclingo\ and \oclingo, respectively,
and its flexibility goes well beyond the   advanced yet still rigid solving processes
of the latter.
\end{abstract}

\section{Introduction}\label{sec:introduction}

Standard Answer Set Programming (ASP; \cite{baral02a}) follows a one-shot process in computing stable models of logic programs.
This view is best reflected by the input/output behavior of monolithic ASP systems like \dlv\ \cite{dlv03a} and \clingo\ \cite{gekakaosscsc11a}.
Internally, however, both follow a fixed two-step process.
First, a grounder generates a (finite) propositional representation of the input program.
Then, a solver computes the stable models of the propositional program.
This rigid process stays unchanged when grounding and solving with separate systems.
In fact, up to now, \clingo\ provided a mere combination of the grounder \gringo\ and the solver \clasp.
Although more elaborate reasoning processes are performed by the extended systems \iclingo\ \cite{gekakaosscth08a} and \oclingo\ \cite{gegrkasc11a} for incremental and reactive reasoning, respectively,
they also follow a pre-defined control loop evading any user control.
%
Beyond this, however,
there is substantial need for specifying flexible reasoning processes,
for instance, when it comes to
interactions with an environment, as in assisted living, robotics, or with users,
advanced search, as in multi-objective optimization, planning, theory solving, or heuristic search,
or recurrent  query answering, as in hardware analysis and testing or stream processing.
Common to all these advanced forms of reasoning is that the problem specification evolves during the reasoning processes,
either because data or constraints are added, deleted, or replaced.

The new \clingo~4 series offers novel high-level constructs for realizing such complex reasoning processes.
This is achieved within a single integrated ASP grounding and solving process
in order to 
avoid redundancies in relaunching grounder and solver programs and 
to benefit from the learning capacities of modern ASP solvers.
To this end,
\clingo~4 
                               complements ASP's declarative input language by control capacities expressed via the embedded scripting languages \lua\ and \python.
On the declarative side, \clingo~4 offers a new directive \lstinline{#program} that allows for structuring logic programs into named and parameterizable subprograms.
The grounding and integration of these subprograms into the solving process is completely modular and fully controllable from the procedural side,
viz.\ the scripting languages embedded via the \lstinline{#script} directive.
For exercising control, the latter benefit from a dedicated \clingo\ library that does not only furnish grounding and solving instructions
but moreover allows for continuously assembling the solver's program in combination with the directive \lstinline{#external}.
Hence, by strictly separating logic and control programs,
\clingo~4 abolishes the need for special-purpose systems for incremental and reactive reasoning, like \iclingo\ and \oclingo, respectively,
and its flexibility goes well beyond the   advanced yet still rigid solving processes
of the latter.



\section{Controlling grounding and solving in \clingo~4}\label{sec:approach}

A key feature, distinguishing \clingo~4 from its predecessors,
is the possibility to structure (non-ground) input rules
into 
subprograms.
To this end,
the directive \lstinline{#program} 
comes with a name and an optional list of parameters.
Once given in the \clingo~4 input,
it gathers all 
rules up to the next such directive (or the end of file)
within a subprogram identified by the supplied name
and parameter list.
As an example,
two subprograms \lstinline{base} and \lstinline{acid(k)}
can be specified as follows:
%
\begin{lstlisting}[xleftmargin=2\parindent]
a(1).
#program acid(k).
b(k).
#program base.
a(2).
\end{lstlisting}
%
%
Note that \lstinline{base}, with an empty parameter list, is a 
dedicated subprogram that,
in addition to rules in the scope of a directive like the one in Line~4,
gathers all rules not preceded by a \lstinline{#program} directive.
Hence, in the above example, 
the \lstinline{base} subprogram
includes the facts \lstinline{a(1)} and \lstinline{a(2)}.
Without further control instructions (see below),
\clingo~4 grounds and solves the \lstinline{base} subprogram only,
essentially yielding the standard behavior of ASP systems.
The processing of
other subprograms, such as \lstinline{acid(k)}
with the schematic fact \lstinline{b(k)},
is subject to scripting control.

For a customized control over grounding and solving,
a \lstinline{main} routine
(taking a control object representing the state of \clingo~4 as argument)
can be specified in either of the embedded scripting languages \lua\ and \python.
For illustration, let us consider two \python\ \lstinline{main} routines: 
\\
\hspace*{2\parindent}%
\begin{minipage}{0.5\linewidth}
\begin{lstlisting}[firstnumber=6]
#script(python)
def main(prg):
    prg.ground("base",[])
    prg.solve()
#end.
\end{lstlisting}  
\end{minipage}
\begin{minipage}{0.5\linewidth}
\begin{lstlisting}[firstnumber=6]
#script(python)
def main(prg):
    prg.ground("acid",[42])
    prg.solve()
#end.
\end{lstlisting}  
\end{minipage}
\\
While the control program on the left matches the default behavior of \clingo~4,
the one on the right ignores all rules in the \lstinline{base} program but rather,
in Line~8, contains a \lstinline{ground} instruction for \lstinline{acid(k)},
where the parameter~\lstinline{k} is instantiated with the term \lstinline{42}.
Accordingly, the schematic fact \lstinline{b(k)} is turned into \lstinline{b(42)},
and the \lstinline{solve} command     in Line~9 yields a stable model consisting of
\lstinline{b(42)} only.
Note that \lstinline{ground} instructions apply to the subprograms
given as arguments,
while \lstinline{solve} triggers reasoning w.r.t.\ all accumulated ground rules.
In fact,
a \lstinline{solve} command makes     \clingo~4     instantiate pending subprograms
and then perform reasoning.
That is,
when Line~9 is replaced, e.g., by \lstinline{print 'Hello!'},
\clingo~4 merely writes out \lstinline{Hello!} but does neither ground any subprogram
nor compute stable models.

In order to accomplish more elaborate reasoning processes,
like those of \iclingo\ and \oclingo\ or customized ones,
it is indispensable to activate or deactivate ground rules on demand.
For instance, former initial or goal state conditions need to be
relaxed or completely replaced when modifying a planning problem, e.g.,
by extending its horizon.
While the     predecessors of \clingo~4 relied on a
\lstinline{#volatile} directive
to provide a rigid mechanism for the expiration of transient rules,
\clingo~4 captures the respective functionalities and customizations
thereof in terms of the directive \lstinline{#external}.
This directive goes back to \lparse\ \cite{lparseManual} and was also
supported by the predecessors of \clingo~4 to exempt (input) atoms
from simplifications fixing them to false.
The \lstinline{#external} directive of \clingo~4 provides a generalization
that, in particular, allows for a flexible handling of yet undefined atoms. 
%

For continuously assembling ground rules evolving at different stages of a
reasoning process, \lstinline{#external} directives declare atoms that may
still be defined by rules added later on.
As detailed in~\cite{gekakasc14a}, such atoms correspond to inputs in terms of module theory~\cite{oikjan06a},
which (unlike undefined output atoms) must not be simplified by fixing their truth value to false.
In order to facilitate the declaration of input atoms, \clingo~4 supports
schematic \lstinline{#external} directives that are instantiated along with
the rules of their respective subprograms.
To this end, a directive like
\begin{lstlisting}[xleftmargin=2\parindent,numbers=none]
#external p(X,Y) : q(X,Z), r(Z,Y).
\end{lstlisting}
is treated similar to the (virtual) rule
\lstinline[xleftmargin=2\parindent,numbers=none]{p(X,Y) :- q(X,Z), r(Z,Y)}
during grounding.
However, the head atoms of resulting ground instances    are merely collected as (external) inputs,
whereas the ground rules as such are discarded.
%
%
Given this, a subprogram        from the \clingo~4 input consists of
all rules within the scope of \lstinline{#program} directives with the
same name and number of parameters, where \lstinline{base} without
parameters is used by default,
along with virtual rules capturing \lstinline{#external} directives
in the same~scope (see~\cite{gekakasc14a} for details).

The instantiation of a subprogram~$\RRG$ with a list $c_1,\dots,c_k$
of parameters, such as \lstinline{acid(k)} above, relies on a list 
$t_1,\dots,t_k$ of terms to replace occurrences of $c_1,\dots,c_k$ with,
both in original rules and virtual rules capturing \lstinline{#external}
directives in~$\RRG$.
The parameter replacement yields a subprogram
$\RRG(c_1/t_1,\dots,c_k/t_k)$, which is instantiated relative to inputs.
For instance, providing the term~\lstinline{42} for parameter~\lstinline{k}
leads to \lstinline{acid(k}$/$\lstinline{42)} consisting of the fact \lstinline{b(42)}.
%
Control instructions guide   the instantiation and assembly of subprograms, 
                                   where \lstinline{ground} instructions issued
before the first or in-between two \lstinline{solve} commands determine rules to
instantiate and join with a module representing the previous state of \clingo~4.

To sum up,
schematic \lstinline{#external} directives are embedded into the grounding process
for a convenient declaration of input atoms from other subprogram instances.
Given that they do not contribute ground rules, but merely qualify (undefined)
atoms that should be exempted from simplifications, \lstinline{#external} directives
address the signature of subprograms' ground instances.
Hence,
it is advisable to condition them by domain predicates%
\footnote{Domain and built-in predicates have unique extensions that
can be evaluated entirely by means of grounding.}
\cite{lparseManual} only,
as this precludes any interferences between signatures and grounder implementations.
As long as input atoms remain undefined, their 
truth values
can be freely picked and modified in-between \lstinline{solve} commands
via \lstinline{assignExternal} instructions,
which thus allow for configuring the inputs to modules
representing \clingo~4 states in order to select among their
stable models.
Unlike that,
the predecessors \iclingo\ and \oclingo\ of \clingo~$4$
always assigned input atoms to false,
so that the addition of rules was necessary to 
accomplish switching truth values. 
However,
for a well-defined semantics,
\clingo~4 like its predecessors builds on the assumption that the
modules induced by subprograms' instantiations are compositional,
which essentially requires definitions of (head) atoms and
mutual positive dependencies to be local to evolving ground programs
(cf.\ \cite{gekakaosscth08a}).
%

\section{Using \clingo~4 in practice}\label{sec:interactive} 

As mentioned above, \clingo~4 fully supersedes its special-purpose predecessors \iclingo\ and \oclingo.
To illustrate this,
we give in Listing~\ref{fig:iclingo:python} a slightly simplified version of \iclingo's control loop in \python.
%
\begin{lstlisting}[float=t,caption={\python\ script implementing \iclingo\ functionality in \clingo\ (\lstinline{iclingo.lp})},label=fig:iclingo:python]
#script(python)
from gringo import Fun, SolveResult

def init(val, default):
    return val if val != None else default

def main(prg):
    stop = str(init(prg.getConst("istop"), "SAT"))
    step = int(init(prg.getConst("iinit"), 0))

    prg.ground("base", [])
    while True:
        step += 1
        prg.ground("cumulative", [step])
        prg.assignExternal(Fun("query", [step]), True)
	print 'STEP {0}'.format(step)
        ret = prg.solve()
        if (stop == "SAT"   and ret == SolveResult.SAT) or \
           (stop == "UNSAT" and ret == SolveResult.UNSAT): break
        prg.releaseExternal(Fun("query", [step]))
#end.
\end{lstlisting}%
%
The full control loop (included in the release) mainly adds handling of further \iclingo\ options.
Roughly speaking, 
\iclingo\ offers a step-oriented, incremental approach to ASP that avoids redundancies by gradually processing the extensions to a problem 
rather than repeatedly re-processing the entire extended problem (as in iterative deepening search).
To this end, a program is partitioned into a 
base part, describing static knowledge independent of the step parameter~\lstinline{t},
a cumulative part, capturing knowledge accumulating with increasing~\lstinline{t},
and 
a volatile part specific for each value of~\lstinline{t}.
These parts were delineated in \iclingo\ by the directives \lstinline{#base}, \lstinline{#cumulative t}, and \lstinline{#volatile t}.
In \clingo~4, all three directives are captured by \lstinline{#program} declarations
along with \lstinline{#external} for volatile rules.

We illustrate this by adapting the Towers of Hanoi encoding from \cite{gekakasc12a} in Figure~\ref{fig:toh}.\captionsetup[subfigure]{justification=raggedright,singlelinecheck=false}
%
\begin{figure}[t]
  \centering
  \subfloat[Towers of Hanoi \rlap{instance}\label{fig:toh:ins}]{
    \begin{minipage}[t]{0.25\textwidth}
\begin{lstlisting}[]
#program base.
peg(a;b;c).
disk(1..4).
init_on(1..4,a).
goal_on(1..4,c).

\end{lstlisting}
    \end{minipage}}\hfill%
  \subfloat[Towers of Hanoi incremental encoding\label{fig:toh:enc}]{
    \begin{minipage}[t]{0.65\textwidth}
\begin{lstlisting}[linerange={1-18}]
#program base.
on(D,P,0) :- init_on(D,P).

#program cumulative(t).
1 { move(D,P,t) : disk(D), peg(P) } 1.

move(D,t) :- move(D,P,t).
on(D,P,t) :- move(D,P,t).
on(D,P,t) :- on(D,P,t-1), not move(D,t).
blocked(D-1,P,t) :- on(D,P,t-1).
blocked(D-1,P,t) :- blocked(D,P,t), disk(D).

:- move(D,P,t), blocked(D-1,P,t).
:- move(D,t), on(D,P,t-1), blocked(D,P,t).
:- disk(D), not 1 { on(D,P,t) } 1.

#external query(t).
:- query(t), goal_on(D,P), not on(D,P,t).
\end{lstlisting}
    \end{minipage}}\negsubfig%
  \caption{Towers of Hanoi instance (\lstinline{tohI.lp}) and incremental encoding (\lstinline{tohE.lp})}
  \label{fig:toh}
\end{figure}
%
The problem instance in Figure~\subref{fig:toh:ins} as well as Line~2 in~\subref{fig:toh:enc} constitute static knowledge and thus belong to the base part.
The transition function is described in the cumulative part in Line~5--15
of Figure~\subref{fig:toh:enc}.
Finally, the query is expressed in Line~18; 
its volatility is realized by making the actual goal condition \lstinline{goal_on(D,P), not on(D,P,t)} subject to the truth assignment to the external atom \lstinline{query(t)}.
%
%
Grounding and solving of the program in Figure~\subref{fig:toh:ins} and~\subref{fig:toh:enc} is controlled by the \python\
script in Listing~\ref{fig:iclingo:python}.
Line~4--9 fix the \lstinline{stop} criterion and initial value of the \lstinline{step} variable.
Both can be supplied as constants \lstinline{istop} and \lstinline{iinit} when invoking \clingo~4.
Once the base part is grounded in Line~11,
the script loops until the \lstinline{stop} criterion is met in Line~18--19.
In each iteration, 
the current value of \lstinline{step} is used 
in Line~14 and~15
to
instantiate 
the subprogram \lstinline{cumulative(t)} 
and to set
the respective
external atom \lstinline{query(t)} to true. 
If the \lstinline{stop} condition is yet unfulfilled
w.r.t.\ the result of solving the extended program,
the current \lstinline{query(t)} atom
is permanently falsified 
(cf.\ Line~17--20), 
thus annulling the corresponding instances of the integrity constraint in Line~18 of Figure~\subref{fig:toh:enc} before they are replaced in the next iteration.


Another innovative feature of \clingo~4 is its incremental optimization.
This allows for adapting objective functions along the evolution of a program at hand.
A simple example is the search for shortest plans when increasing the horizon in non-consecutive steps.
To see this,
recall that literals in minimize statements (and analogously weak constraints) are supplied with a sequence of terms of the form $w@p,\vec{t}$,
where $w$ and $p$ are integers providing a weight and a priority level and $\vec{t}$ is a sequence of terms (cf.\ \cite{ASPCore2}).
As an example, consider the subprogram:
\begin{lstlisting}[numbers=none,escapechar=?,xleftmargin=2\parindent]
#program cumulativeObjective(t).
#minimize{ W@P,X,Y,t : move(X,Y,W,P,t) }.
?\textit{\% or}? :~ move(X,Y,W,P,t). [W@P,X,Y,t]
\end{lstlisting}
When grounding and solving \lstinline{cumulativeObjective(t)} for successive values of \lstinline{t},
the solver's objective function (per priority level \lstinline{P}) is 
gradually extended 
with new atoms over \lstinline{move/5}, 
and   all previous ones are kept.

Moreover, for enabling the removal of literals from objective functions,
we can use externals:
\begin{lstlisting}[numbers=none,xleftmargin=2\parindent]
#program volatileObjective(t).
#external activateObjective(t).
#minimize{ W@P,X,Y,t : move(X,Y,W,P,t), activateObjective(t) }.
\end{lstlisting}  
The subprogram \lstinline{volatileObjective(t)} behaves like \lstinline{cumulativeObjective(t)} as long as the external atom \lstinline{activateObjective(t)} is true.
Once it is set to false, all atoms over \lstinline{move/5} with the
corresponding 
term for~\lstinline{t} 
are dismissed from 
objective functions.


A reasoning process in \clingo~4 is partitioned into a sequence of solver invocations. 
We have seen how easily the solver's logic program can be altered at each step.
Sometimes it is useful to do this in view of a 
previously obtained stable model.
For this purpose, 
the \lstinline{solve} command can be equipped with an (optional) callback
function \lstinline{onModel}. 
For each stable model found during a call to \lstinline{solve(onModel)},
an object encompassing the model is passed to 
\lstinline{onModel},
whose implementation can then 
access and inspect the model.
A typical example is the addition of constraints based on the last model that are then supplied to the solver before computing the next one.
An application is theory solving by passing (parts of) the last model to a theory solver for theory-based consistency checking or for providing
the value of an externally 
evaluated objective function.
%
Moreover,
\clingo~4 also furnishes an asynchronous solving function \lstinline{asolve} 
that launches an interruptable solving process in the background.
%
This is particularly useful in reactive settings 
in order to stop solving
upon the arrival of new external information.

Similarly, the configuration of \clasp\ can be changed at each step via the function \lstinline{setConf}, 
taking a string including command line options along with a flag indicating whether the previous configuration is updated or replaced as arguments.
For instance, this allows for changing search parameters, reasoning modes, number of threads, etc.
Changing search parameters is of interest when addressing 
computational tasks involving the generation of several models,
like optimal planning, multi-criteria optimization, or heuristic search.
Apart from analyzing the previous model via the \lstinline{onModel} callback,
one 
can also monitor the search progress by means of the function \lstinline{getStats},
returning an     object encapsulating up to 135 attributes of the previous search process.
Furthermore,
\clingo~4 allows for customizing 
the heuristic values of variables, as described in \cite{gekaotroscwa13a}.
At a higher level, a user may simply want to explore the set of models, and decide to compute first one, then all, and then the intersection or union of all models. 
This can be interleaved with the addition of 
subprograms 
via the function \lstinline{add}, 
which may in turn include \lstinline{#external} directives to declare
temporary hypotheses.
The experienced reader may note that this can be done fully interactively by 
means of 
I\python. 
Practical
examples for 
the mentioned features can be found in the releases 
at \cite{potassco}.



\section{Related work}
\label{sec:related:work}

Although 
\clingo~3 \cite{gekakosc11a} already featured \lua\ as an embedded scripting language, 
its usage was limited to 
(deterministic) computations during grounding;
neither were library functions furnished by \clingo~3.

Of particular interest is
\dlvhex\ \cite{figeiaresc13a}, an ASP system aiming at the integration of external computation sources.
For this purpose, \dlvhex\ relies on higher-order logic programs using external higher-order atoms for software interoperability.
Such external atoms should not be confused with \clingo's \lstinline{#external} directive because they are evaluated via 
procedural means during solving.
Given this,
\dlvhex\ can be seen as an \emph{ASP modulo Theory} solver, similar to SAT modulo Theory solvers \cite{niolti06a}.
In fact, \dlvhex\ uses \gringo\ and \clasp\ as back-ends and follows the design of the \emph{ASP modulo CSP} solver \clingcon\ \cite{ostsch12a}
in communicating with 
external ``oracles'' through \clasp's 
post propagation mechanism.
In this way, 
theory solvers are tightly
integrated into the ASP system and 
have access to the solver's partial assignments.
Unlike this, the light-weighted theory solving approach offered by \clingo~4 can only provide access to total (stable) assignments.
It is thus interesting future work to investigate in how far \dlvhex\ can benefit from lifting its current low-level integration into \clasp\ to a
higher level in combination with \clingo~4.
Clearly, the above considerations also apply to extensions of \dlvhex, such as \acthex\ \cite{figeiaresc13a}.
%
Furthermore,
\jdlv\ \cite{felegrri12a} 
encapsulates the \dlv\ system to
facilitate
one-shot ASP solving 
in \java\ environments by providing means to 
generate and process logic programs,
and to afterwards extract their stable models.

The procedural attachment to the \idp\ system \cite{powide11a,debobrde14a}
builds on interfaces to \cpp\ and \lua.
Like \clingo~4, it allows for 
evaluating functions during grounding,
calling the grounder and solver multiple times,
inspecting solutions, and
reacting to external input after search.
%
The emphasis, however, lies 
on 
high-level control 
blending in with \idp's modeling language,
while
\clingo~4 offers more fine-grained control over the grounding and solving process,
particularly 
aiming at a flexible incremental assembly of programs from subprograms.

In SAT, incremental solver interfaces from low-level APIs are common practice.
Pioneering work was done in \minisat\
\cite{eensor03a}, 
furnishing a \cpp\ interface for solving under assumptions. 
In fact, the \clasp\ library underlying \clingo~4 builds upon this functionality to implement incremental search 
(see \cite{gekakaosscth08a}).
Given that SAT deals with propositional formulas only,
solvers and their APIs lack 
support for modeling languages and grounding.
Unlike this,
the SAT modulo Theory solver \ZZZ\ \cite{dembjo08a}
comes with a \python\ API that,
similar to \clingo~4, 
provides a library for controlling 
the solver as well as 
language bindings for constraint handling.
In this way, \python\ can be 
used as a modeling language for \ZZZ.


\section{Discussion}\label{sec:discussion}

The new \clingo~4 system complements ASP's declarative input language by control capacities expressed by embedded scripting languages.
This is accomplished within a single integrated ASP grounding and solving process in which a logic program may evolve over time.
The addition, deletion, and replacement of programs is controlled procedurally by means of \clingo's dedicated library.
The incentives for evolving a logic program are manifold and cannot be captured with the standard one-shot approach of ASP.
Examples include
unrolling a transition function, as in planning,
interacting with an environment, as in assisted living, robotics, or stream reasoning,
interacting with a user exploring a domain,
theory solving,
and advanced forms of search.
Addressing these demands
                by embedded scripting languages provides us with a generic and transparent approach.
Unlike this, previous systems, like \iclingo\ and \oclingo, had a dedicated purpose involving rigid control capacities buried in monolithic programs.
Rather than that,
the basic technology of \clingo~4 allows us to instantiate subprograms in-between solver invocations in a fully customizable way.
On the declarative side,
the availability of program parameters and the embedding of
\lstinline{#external} directives  
into the grounding process provide 
great flexibility in modeling schematic subprograms.
In addition, the possibility of assigning input atoms facilitates
the implementation of applications such as query answering or sliding window
reasoning, as truth values can now be switched without manipulating a logic program.

The semantic underpinnings of our framework in terms of module theory 
capture
the dynamic combination of logic programs in a generic way.
It is interesting future work to investigate how dedicated change operations
whose interest was so far mainly theoretic, like 
updating \cite{alpeprpr02a} 
or forgetting \cite{zhafoo06a}, 
can be put into practice within this framework.

The input language of
\clingo~4 extends the 
\textit{ASP-Core-2} standard \cite{ASPCore2}.
Although we have presented \clingo~4 for normal logic programs,
we mention that it accepts (extended) disjunctive logic programs, 
processed via 
the multi-threaded solving approach described in \cite{gekasc13a}.
In version 4.3, \clingo\ moreover 
embeds \clasp~3, featuring
domain-specific heuristics \cite{gekaotroscwa13a}
and
optimization using unsatisfiable cores \cite{ankamasc12a}.
\clingo~4 is freely available at \cite{potassco}, 
and its releases include many best practice examples
illustrating the aforementioned application scenarios.


\paragraph{Acknowledgments}

This work was partially funded by DFG grant SCHA 550/9-1.




\end{document}